\RequirePackage{fix-cm}
\documentclass[10pt,a4paper]{article}
\usepackage{graphicx}

\usepackage[pdftex,dvipsnames]{xcolor} 
\usepackage[algo2e]{algorithm2e}
\usepackage{algorithm}
\usepackage{algorithmic}
\usepackage{array}
\usepackage{amssymb}
\usepackage{amsmath}
\usepackage{amsthm}
\usepackage{amsfonts}
\usepackage{empheq}
\usepackage{enumerate}
\usepackage{etex}
\usepackage[export]{adjustbox}
\usepackage{graphicx}
\usepackage{float}
\usepackage{makeidx} 
\usepackage{tikz}
\usepackage{mathabx}
\usepackage{tabulary}
\usepackage{tabularx}
\usepackage{color}
\usepackage{xargs}                      
\usepackage{fancyhdr}
\usepackage{url}
\usepackage{blindtext}
\usepackage{fixltx2e}
\usepackage{caption}

\usepackage{regexpatch}
\usepackage[english]{babel}
\usepackage[normalem]{ulem}

\theoremstyle{definition}
\newtheorem{exmp}{Example}[section]

\usepackage{footmisc}
\usepackage[colorinlistoftodos,prependcaption,textsize=small]{todonotes}

\newcommandx{\unsure}[2][1=]{\todo[linecolor=red,backgroundcolor=red!25,bordercolor=red,#1]{#2}}
\newcommandx{\change}[2][1=]{\todo[linecolor=blue,backgroundcolor=blue!25,bordercolor=blue,#1]{#2}}
\newcommandx{\info}[2][1=]{\todo[linecolor=OliveGreen,backgroundcolor=OliveGreen!25,bordercolor=OliveGreen,#1]{#2}}
\newcommandx{\improvement}[2][1=]{\todo[linecolor=Plum,backgroundcolor=Plum!25,bordercolor=Plum,#1]{#2}}
\newcommandx{\thiswillnotshow}[2][1=]{\todo[disable,#1]{#2}}

\newcommand*\circled[1]{\tikz[baseline=(char.base)]{
            \node[shape=circle,draw,fill=black, inner sep=0.5pt] (char) {#1};} }

\usepackage{pgf-umlsd}
\usetikzlibrary{arrows, shadows, positioning}
\usepackage{makeidx}  

\usepackage{fancybox}
\usepackage{wrapfig} 
\usepackage{nameref}
\usepackage{fixltx2e}

\newcommand{\amess}[7][0]{
  \stepcounter{seqlevel}
  \path
  (#2)+(0,-\theseqlevel*\unitfactor-0.7*\unitfactor) node (mess from) {};
  \addtocounter{seqlevel}{#1}
  \path
  (#4)+(0,-\theseqlevel*\unitfactor-0.7*\unitfactor) node (mess to) {};
  \draw[->,>=angle 60] (mess from) -- (mess to) node[midway, below]
  {#3};

  \if R#5
    \node (#3 from) at (mess from) {\llap{#6~}};
    \node (#3 to) at (mess to) {\rlap{~#7}};
  \else\if L#5
         \node (#3 from) at (mess from) {\rlap{~#6}};
         \node (#3 to) at (mess to) {\llap{#7~}};
       \else
         \node (#3 from) at (mess from) {#6};
         \node (#3 to) at (mess to) {#7};
       \fi
  \fi
}

\renewcommand{\mess}[4][0]{
  \stepcounter{seqlevel}
  \path
  (#2)+(0,-\theseqlevel*\unitfactor-0.7*\unitfactor) node (mess from) {};
  \addtocounter{seqlevel}{#1}
  \path
  (#4)+(0,-\theseqlevel*\unitfactor-0.7*\unitfactor) node (mess to) {};
  \draw[->,>=angle 60] (mess from) -- (mess to) node[midway, above]
  {#3};

  \node (\detokenize{#3} from) at (mess from) {};
  \node (\detokenize{#3} to) at (mess to) {};
}

\tikzset{every picture/.append style={scale=0.9}}

\usepackage{listings}
\usepackage{parcolumns}

\usepackage{color}
\definecolor{gray}{rgb}{0.4,0.4,0.4}
\definecolor{darkblue}{rgb}{0.0,0.0,0.6}
\definecolor{cyan}{rgb}{0.0,0.6,0.6}

\definecolor{dkgreen}{rgb}{0,0.6,0}
\definecolor{mauve}{rgb}{0.58,0,0.82}

\lstdefinestyle{smallxml}{
  language=xml,
  numbers=left,
  stepnumber=1,
  numbersep=10pt,
  tabsize=3,
  showspaces=false,
  showstringspaces=false
}

\begin{document}

\title{SDN Access Control for the Masses}


%
%

\author{Nicolae Paladi$^1$ \and Christian Gehrmann$^2$}
\date{%
    $^1$RISE SICS\\%
    $^2$Lund University\\[2ex]%
}


\maketitle
\begin{abstract}
The evolution of \textit{Software-Defined Networking} (SDN) has so far been predominantly geared towards defining and refining the abstractions on the forwarding and control planes.
However, despite a maturing south-bound interface and a range of proposed network operating systems, the network management application layer is yet to be specified and standardized.
It has currently poorly defined access control mechanisms that could be exposed to network applications.
Available mechanisms allow only rudimentary control and lack procedures to partition resource access across multiple dimensions.

We address this by extending the SDN north-bound interface to provide control over shared resources to key stakeholders of network infrastructure: network providers, operators and application developers.
We introduce a taxonomy of SDN access models, describe a comprehensive design for SDN access control and implement the proposed solution as an extension of the ONOS network controller intent framework.

\end{abstract}

\section{Introduction}
\label{sec:Intro}

In recent years, research focus on software-defined networking (SDN) shifted from the maturing forwarding plane abstraction~\cite{doria:2010,pfaff:2015} and corresponding south-bound application programming interfaces (APIs) \cite{halpern:2010,mckeown:2008}, to richer network control, management and functionality, as well as initial attempts to define a north-bound API.
However, the field has advanced unevenly.
Currently available north-bound APIs introduce new security risks~\cite{klaedtke:2014}, and the lack of \textit{network resource access control} is one vivid example of this.
Security research in SDN has focused on managing access using topology-specific, low-level resources -- such as switch ports and fine-grained bit-matching of packet flows.
While this approach is valid and necessary, it does not meaningfully address the needs of either network application \textit{developers} (who create applications operating with network functionality on a higher level of abstraction) or network \textit{operators} (who must consider the security and resource management implications of deploying such applications).
As a result, developers face a choice between two undesirable alternatives: assume during development that applications have full, exclusive, and continuous access to deployment resources, or develop custom applications with in-built awareness of the network topology.
However, development perils do not end there: in either case network applications operate with \textit{low-level} device resources, mostly unfamiliar to system developers.
Finally, at deployment time, operators lack an overview of the resource access granted to network applications once they have been deployed.

Recent efforts aim to provide a declarative paradigm for implementation - independent interaction between network service consumers and providers~\cite{nbi:2016}.
However, such efforts do not currently include any access control mechanisms beyond specifying service resource constraints, mostly aimed towards satisfying Quality-Of-Service network provider policies.
Several important prior contributions to SDN access control~\cite{ferguson:2012,monaco:2013,klaedtke:2014} do not provide the necessary abstraction level to allow exposing access control functionality to potentially malicious SDN applications.
We address the above issues by introducing a novel taxonomy of access models for SDN infrastructure resources and by describing and implementing a North-bound Access Control API ($\mathsf{NACA}$) enforcement mechanism for SDN deployments.

\textit{For developers}, $\mathsf{NACA}$ brings simple, clear and usable tools to declare resource requirements of their network applications.

\textit{For operators}, the API provides the tools to
(1) obtain an overview of the access to network resources provided to the applications deployed on the SDN infrastructure and
(2) assess the security implications of deploying network applications, considering the resource access they require.
This fills the gap in the available tools for managing access control to device resources in SDN deployments and helps answer common questions -- such as ``Which applications can read the complete topology of the network?'', or ``Which applications can modify network flows and in which ways?'' -- that are currently difficult to answer in a given SDN deployment;
(3) limit the extent of application access to network resources through resource-specific policies.

To implement the API on the control plane\footnote{In the remainder of this paper, we use the terms \textit{forwarding plane, control plane, management plane, application plane} as defined in~\cite{haleplidis:2015}.}, we build our approach on best practices from the fields of operating system security and programming language security -- such as 
code signing for origin verification; 
code integrity verification; 
capability-based access control,
use of a reference monitor, etc. -- to enforce access restrictions to SDN infrastructure resources.
Furthermore, we propose leveraging recent developments in execution isolation in order to ensure the robustness of the deployment in the face of a powerful adversary.

In a nutshell, the proposed approach is as follows.
We adopt the common access control convention and identify two types of entities: \textit{subjects} -- network applications\footnote{And implicitly their users}, often referred to as ``Virtual Network Function'' (VNFs), and \textit{objects} -- network resources.
$\mathsf{NACA}$ \textit{access masks} are policies that describe the objects that a certain subject can access, as well as the types of and constraints on actions that can be performed on the object attributes.
Examples of~\textit{types} of actions on objects are reading statistics, modifying configuration state, or subscribing to notifications.
Examples of~\textit{object attributes} are geographical placement of objects, temporal limitations, execution environment visibility, etc.
Resource access \textit{requirements} declared by a subject are combined with operator policies to define a distinct immutable \textit{access mask} that persists throughout the lifetime of subject instances.
Next, we reliably \textit{tag} requests issued by subjects with their assigned access mask and use a \textit{reference monitor} to ensure that access to the \textit{network information base} is only granted to requests satisfying the constraints of the access mask. 
In contrast to previous work in this area, $\mathsf{NACA}$ does not depend on the use of a particular south-bound-API.
It operates entirely on the north-bound API and the implementation on the controller plane.
Our contribution is as follows:
\begin{itemize}
	\item We introduce a taxonomy of access models for network infrastructure resources;
	\item We introduce a North-bound Access Control API for network infrastructure operators and application developers;
	\item We propose a novel access control enforcement mechanism for network resources in the software-defined networking model;
	\item We describe an implementation of the proposed API on the control plane;
	\item We demonstrate the feasibility of the solution through an integration with an open source SDN controller platform. 
\end{itemize}

The remainder of the paper is organized as follows:
in Section~\ref{sec:sys_adversarial_model} we introduce the system and threat model;
next, in Section~\ref{sec:NACA} we describe the north-bound access control API and its internals, 
followed in Section~\ref{sec:implementation_onos} by a detailed description of the implementation with the ONOS SDN platform and the evaluation results in Section~\ref{sec:evaluation}.
We review the related work in Section~\ref{sec:related_work}, outline future work in Section~\ref{sec:future_work} and conclude in Section~\ref{sec:conclusion}.


\section{System and Threat model}
\label{sec:sys_adversarial_model}

\subsection{System model}
\label{subsec:sys_model}

\begin{figure}[t]
\centering
\includegraphics[width=0.6\textwidth]{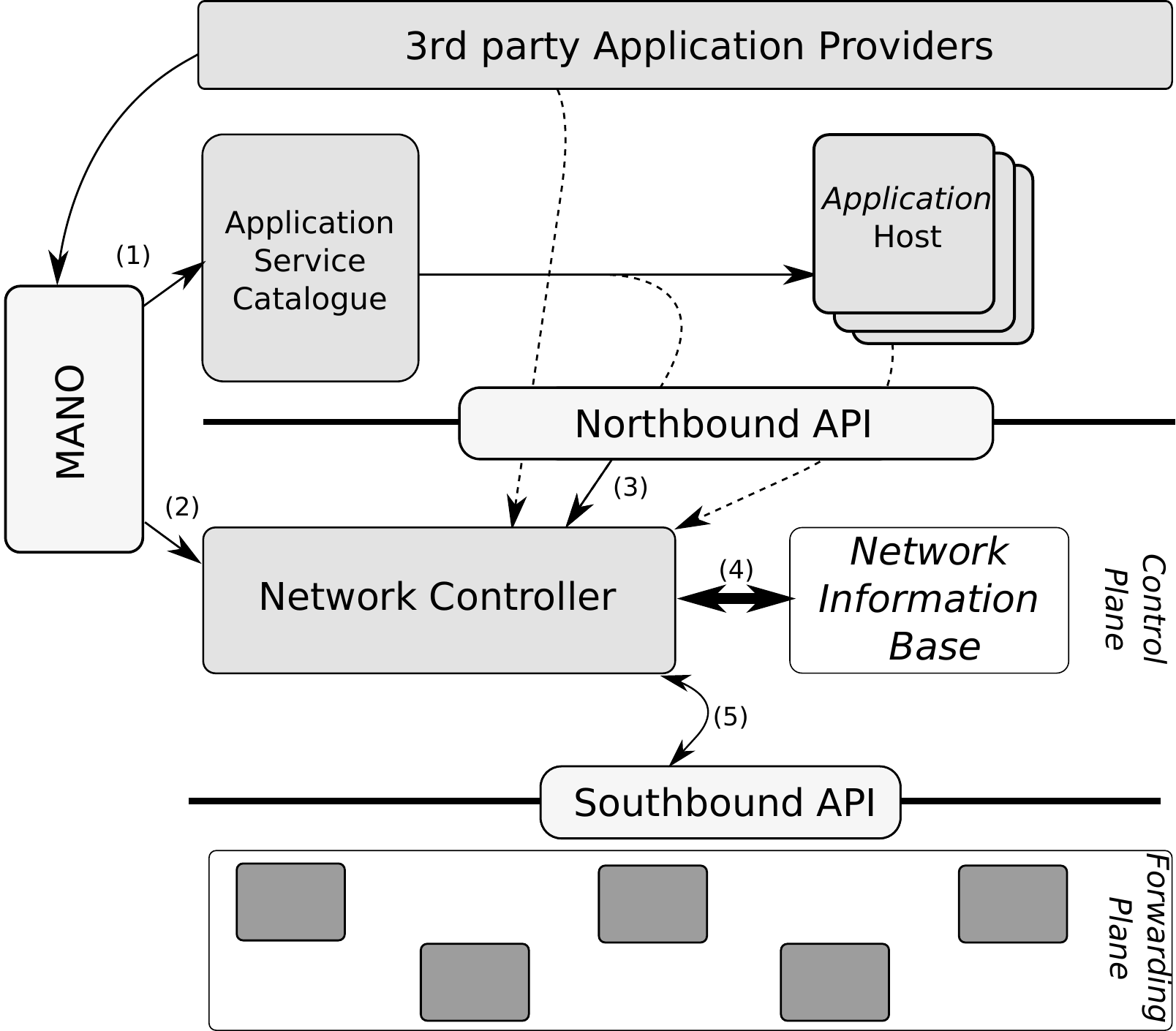}
\caption{\small The SDN architectural model: 
(1) populate service catalogue; 
(2) deploy applications; 
(3) applications interact with network controller and (4) query the NIB; 
(5) Network controller performs configuration actions on forwarding plane.}
\label{fig:SDN_architecture}
\end{figure}

The software-defined networking model aims to separate the network \textit{forwarding plane} -- i.e. the collection of network devices responsible for forwarding traffic -- from the control plane -- i.e. a collection of functions controlling network devices, defining the network topology and network connectivity policies~\cite{haleplidis:2015,casado:2014}.
Figure~\ref{fig:SDN_architecture} illustrates a high-level architecture of the software-defined networking model.

The forwarding plane includes hardware and software \textit{switches}.
Early SDN models envisioned switches that are optimized for forwarding performance, lack decision logic and only forward packets matching \textit{flow definitions} -- i.e. packet forwarding rules -- in their \textit{forwarding information base}~\cite{nadeau:2013}.
Later contributions delegate more functional responsibility to switches, while maintaining the capability to selectively upstream packets (or packet data) to controllers~\cite{bifulco:2016}.
Mismatching packets are discarded or redirected to the \textit{control plane} through the \textit{south-bound API} -- a set of vendor-agnostic instructions for communication between forwarding and control planes; 
this API is often limited to flow-based \textit{traffic control} of the forwarding plane, while \textit{management} of the forwarding plane is done through a configuration database~\cite{pfaff:2015}.

On the \textit{control plane}, network operator goals are translated into discrete routing policies based on the \textit{global network view}, e.g. a graph representation of the network topology.
A core element is the \textit{network controller} -- a logically centralized component that manages network communication in a deployment by updating the FIB with specific forwarding rules.
The network controller compiles forwarding rules based on three inputs: the (dynamic) global network view, the configuration goals of the network operator, and the output of the installed \textit{network applications}.
The network view built by the controller is maintained in a \textit{network information base} (NIB)~\cite{koponen:2010}.
This may either include the entire network topology or a \textit{slice} of it (e.g. in multi-tenant deployments).

\begin{figure}[t]
\centering
\includegraphics[width=\textwidth]{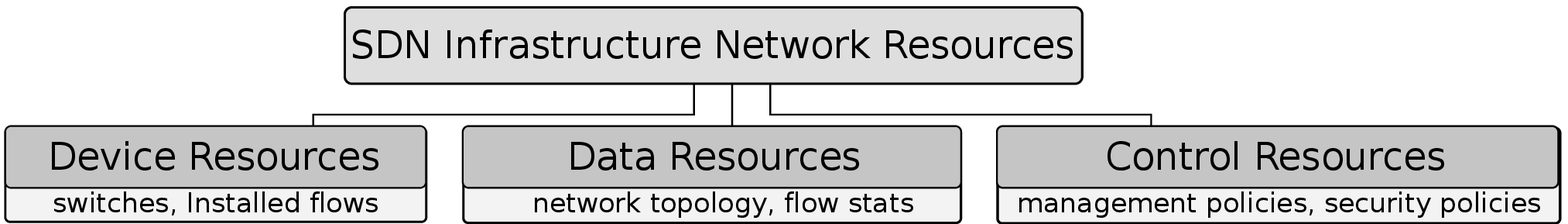}
\caption{\small SDN infrastructure network resources.}
\label{fig:resources}
\end{figure}

The NIB describes all resources of the SDN deployment reachable by the controller.
\textit{We use a broad definition of the term resources}, to encompass software components used to achieve network communication goals (e.g. virtual or physical switches), \textit{information} about such network components, and  interactions involving them.
From a network application point of view, we distinguish three resource categories:
\textit{device resources}, e.g. forwarding plane components; 
\textit{data resources}, e.g. network topology, flow statistics, forwarding logic; 
and \textit{control resources}, e.g. management policies (Figure~\ref{fig:resources}).

Operators use network management applications to implement network functionality using high-level commands.
Network applications -- also known as ``middleboxes'' -- often appear as hardware components in traditional networks; 
however, alternatives such as VNFs -- e.g. software implementations of firewalls, traffic shapers, etc. -- are better suited for dynamic SDN deployments and hence are becoming increasingly popular.
Applications communicate with the network controller and are used for network management, based on operator-defined policies and network state.

There is currently no single widely adopted interface between applications and network controllers (i.e. a ``north-bound API'').
Multiple distinct implementation-specific interfaces are used by network controllers~\cite{onos:2016,opendaylight:2016}.
We distinguish three emerging network application deployment models:
\begin{enumerate}
	\item \textit{Locally installed applications}, developed in-house or deployed through e.g. ``SDN App Stores'' \cite{scott:2015b}.
	\item \textit{Managed applications}, operating in an “Software-as-a-Service” model, i.e. on the premises of a network function provider~\cite{matthews:2010,etsi:2016}.
	\item \textit{Hybrid applications}, where a back-end executing on the application provider premises interacts with a front-end on the network provider infrastructure.
\end{enumerate}
To facilitate function isolation, scalability and deployment flexibility, applications are commonly deployed as virtualized components, in e.g. virtual machines or containers.
We define \textit{candidate applications} as the applications available for deployment from the service catalogue, which can contain both complete application images for locally installed applications or configuration definitions in the managed applications. 

A \textit{management and network orchestration} (MANO) component monitors the SDN infrastructure and takes actions to ensure availability and satisfy performance requirements.
Such actions include component creation, deployment, migration and destruction.
A candidate application becomes an \textit{installed application} once it is granted \textit{access} to the SDN resources described in the NIB.

We define \textit{access by an application to SDN resources} as the capability to execute commands on device resources or modify their state; 
create, read or write control resources and data resources.
Applications exercise access through \textit{requests} submitted to the controller over the north-bound API; 
the requests are further compiled into a limited number of queries by the network controller and submitted to the network information base.
A network controller can only issue queries following a corresponding request by a network application.

Finally, we assume that all communication between network elements is encrypted and authenticated.
The MANO component provisions all necessary encryption and authentication credentials to the network elements at deployment time.

\subsection{Threat model}
\label{subsec:adv_model}
\begin{figure}[t]
\centering
\includegraphics[width=0.8\textwidth]{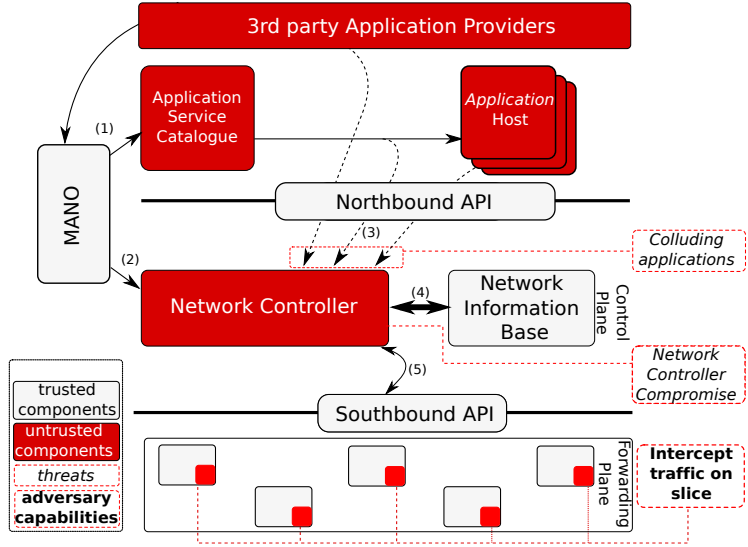}
\caption{\small Adversary capabilities}
\label{fig:adv-model}
\end{figure}
We next describe the threat model (illustrated in Figure~\ref{fig:adv-model}), along with core security assumptions on which we base our design.
The adversary (\textit{Adv}) controls the applications installed on the network slice and can request access to arbitrary device resources.
Furthermore, the \textit{Adv} can collude several applications to achieve a defined purpose, e.g. take over the device resources allocated to benign applications (provoking a Denial-of-Service attack on them).
It can intercept, record, forge, drop and replay any message on its network slice and is only limited by the constraints of the employed cryptographic methods.
Furthermore, it can analyze network traffic patterns through passive probing and may disrupt or degrade network connectivity to achieve its goals.
The adversary can craft malicious packets to exploit vulnerabilities in the request processing functionality of the network controller: an adversary may use a network application to submit malicious requests to trigger escalation of resource access permissions for the respective application.
However, the \textit{Adv} can only interact with the NIB through queries produced by the network controller, based on requests issued by installed network applications.
Briefly, the capabilities of the adversary are similar to the ones of a malicious application provider or operator, whose applications are installed on a network slice in a ``Network-as-a-Service'' provisioning model.

\subsection{Notation and Cryptographic Primitives}
\label{subsec:cryptoPrimitives}
The set of all binary strings of length $n$ is denoted by $\left\{0,1\right\}^n$, and the set of all finite binary strings as $\left\{0,1\right\}^*$.
Given a set $U$, we refer to the $i^{th}$ element as $u_i$.
Additionally, we use the following notations for cryptographic operations:
\begin{itemize}
  \item For an arbitrary message $m \in \left\{0,1\right\}^*$, we denote by $c = \mathsf{Enc}\left(K, m\right)$ a symmetric encryption of $m$ using the secret key $K \in \left\{0,1\right\}^*$. The corresponding symmetric decryption operation is denoted by $m = \mathsf{Dec}(K,c)=\mathsf{Dec}(K, \mathsf{Enc}(K,m))$.
  \item We denote by $\mathsf{pk/sk}$ a public/private key pair for a public key encryption scheme. Encryption of message $m$ under the public key $\mathsf{pk}$ is denoted by $c = \mathsf{Enc_{pk}}\left(m\right)$\footnote{Alternative notation used for clarity is $\left\{m\right\}_{\mathsf{pk}}$.} and the corresponding decryption operation by $m = \mathsf{Dec_{sk}}(c)=\mathsf{Dec_{sk}}(\mathsf{Enc_{pk}}(m))$.
  \item A digital signature over a message $m$ is denoted by $\sigma = \mathsf{Sign_{\mathsf{sk}}}(m)$. The corresponding verification operation for a digital signature is denoted by ${b =\mathsf{Verify}_{\mathsf{pk}}(m,\sigma)}$, where $b = 1$ if the signature is valid and $b=0$ otherwise.
  \item A Message Authentication Code ($\mathsf{MAC}$) using a secret key $K$ over a message $m$ is denoted by ${\mu = \mathsf{MAC}(K,m)}$.
\end{itemize}


\section{Taking Control Over Network Resources}
\label{sec:NACA}

Decoupling abstraction layers is a core benefit of the SDN model.
It allows to combine solutions from distinct providers across the abstraction layers of a network infrastructure while maintaining encapsulation.
This also applies to the interface between the application layer and the rest of the SDN deployment: 
on the one hand, application developers are often oblivious to packet switching details and network functionality internals.
On the other hand, operators may want to withhold details of their SDN deployments from potentially malicious applications and only allow them to interact with the SDN infrastructure through a restricted policy interface.

\subsection{Access Classification Scheme}
\label{subsec:ACS}

\begin{figure*}[b]
\centering
\includegraphics[width=0.8\textwidth]{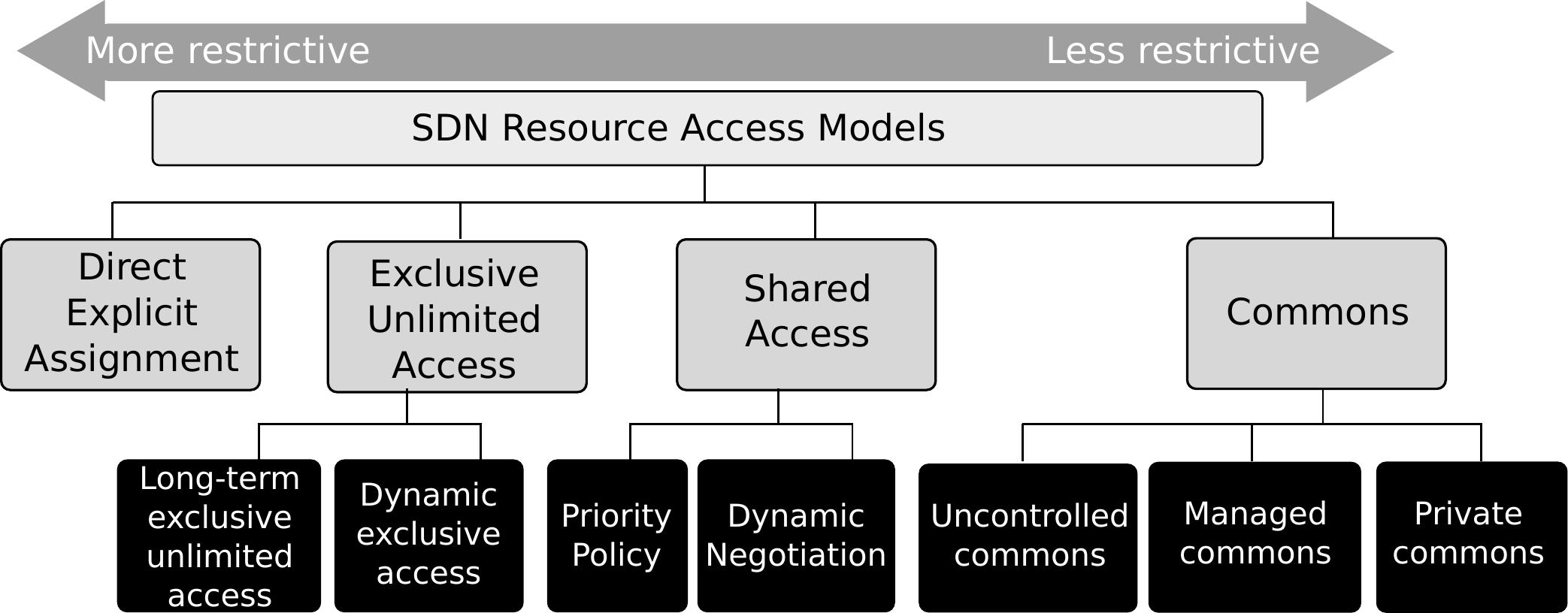}
\caption{\small Taxonomy of SDN resource access models.}
\label{fig:accClass}
\end{figure*}

Network applications require a variety of network resources to fulfill their functional requirements.
In some cases, they may require temporary \textit{exclusive} access, e.g. for atomic updates to the NIB~\cite{schiff:2016}.
However, they seldom -- if ever -- require \textit{complete} and \textit{indefinite} access to \textit{all} resources.
Furthermore, operators may wish to limit the access of applications to SDN resources.
For example, a passive network intrusion detection function may only need ``read'' access to device resources, control resources and data resources, but it need not be able to modify the network state; 
in addition, the network operator may intend to limit the access of the intrusion-detection application to traffic from endpoints located in a certain jurisdiction.
Similar to other domains where multiple parties access distributed resources (e.g. radio spectrum access~\cite{buddhikot:2007}), we foresee multiple models of managing access to SDN resources.
Based on an extensive review of existing literature on software-defined networking~\cite{sherwood:2010,koponen:2010,nadeau:2013,scott:2015,schiff:2016}, as well as earlier classifications of access to resources in distributed systems~\cite{buddhikot:2007,casavant:1988}, we next propose a taxonomy of SDN resource access models (see Figure~\ref{fig:accClass}).
While this scheme covers the access models and use cases identified in the reviewed literature, it can be extended with novel and emerging access models in the future.

\subsubsection{Direct Explicit Assignment}
In this model, the operator explicitly prescribes the SDN resources each application can access, the actions that can be performed on the resources, the duration of the granted access and other (potentially resource-specific) attributes.
This approach is suitable in cases when applications must have guaranteed access to certain resources or vice-versa, when particularly limited or security-sensitive SDN resources must be explicitly assigned to certain applications to be accessible.
Direct explicit assignment might not be applicable to either \textit{all} applications or SDN resources in the deployment.
The direct assignment model is the most restrictive in the proposed taxonomy.

\subsubsection{Exclusive Unlimited Access}
In this model, the operator allows the application exclusive, unlimited access to listed SDN resources under defined constraints -- e.g. geographical placement, jurisdiction, duration, etc;
it encompasses two variants:
\begin{itemize}
	\item \textbf{Long-term exclusive unlimited access:} an installed application \textit{A} has unfettered access to SDN resources until its termination and cannot delegate its access permissions to a different application;
	\item \textbf{Dynamic exclusive access:} the installed application \textit{A} has unfettered access to SDN resources.
However it can delegate its permissions to other installed applications (e.g. \textit{B} and \textit{C}).
All permissions for applications \textit{A, B, C} are revoked once application \textit{A} is terminated.
\end{itemize}

\subsubsection{Shared Access}
In this model, the access to SDN resources allocated to an installed application \textit{A} is \textit{shared} with one or more installed applications (e.g. \textit{B} and \textit{C}).
This model is further detailed into the following two variants:
\begin{itemize}
	\item \textbf{Priority policy:} in this case, if applications \textit{B} and \textit{C} may attempt to access the SDN resources allocated to \textit{A}, their requests will \textit{always} be denied in case of a conflicting request from \textit{A}.
	\item \textbf{Dynamic negotiation:} application \textit{A} \textit{may} issue requests with varying priorities when prompted by other installed applications \textit{B} and \textit{C}.
In this case, higher-priority requests from applications \textit{B} and \textit{C} would be accepted.
Implementation of this variant may involve intricate details on dynamic access negotiation between applications.
\end{itemize}

\subsubsection{Commons}
The network operator may apply this \textit{least} restrictive model to applications which require access to the same pool of SDN resources and have the same trust level.
This model contains three variants:
\begin{itemize}
\item \textbf{Uncontrolled commons:} installed applications compete unrestricted for access to the allocated SDN resources.
Conflict resolution mechanisms (as proposed in \cite{ferguson:2013,klaedtke:2014}) can be used to prioritize conflicting requests.
Lack of effective conflict resolution can impair the functionality of this variant.
\item \textbf{Managed commons:} installed applications compete for access to the allocated SDN resources; 
in case of request conflicts, applications negotiate access using peer-to-peer protocols.
Compared to the uncontrolled commons variant, this reduces the conflict resolution overhead, at the cost of increased communication between installed applications.
\item \textbf{Private commons:} this variant includes elements of the \textit{dynamic exclusive access} variant described above.
Installed application \textit{A} with exclusive unlimited access to a set of SDN resources (and delegation permissions) may allow access for other applications using one of the commons variants described above.
\end{itemize}

Once one or more SDN resource access models have been selected, the resource access control is expressed through policies, as described below.

\subsection{Policies for infrastructure management}
\label{subsec:policies-infram}

Kephart et al~\cite{kephart:2004} describe three types of infrastructure management policies: \textit{goal policies}, \textit{utility function policies} and \textit{action policies}.
Goal policies and utility function policies are most suitable to specify enterprise business objectives and Service Level Agreements.
In a datacenter scenario, a goal policy example may be \textit{``Response time of Gold Class should be less than 100ms''}, while a utility function policy
can be \textit{``Maximize the sum of Gold and Silver Classes''}.
In practice, such high-level commands are transformed into action policies, often in the form of 
`\texttt{IF (Condition) THEN (Action)}', e.g. `\texttt{IF (Gold\_Class.Response\_Time > 100 ms) THEN (increase CPU by 5\%)}'. 
Considering the widely adopted SDN south-bound API protocols, it is clear that they are unsuitable for expressing goal policies or utility function policies;
it is equally clear that formulating action policies requires intimate knowledge of the SDN deployment.
Instead, functional details of SDN deployments can be encapsulated and selectively exposed to network applications through a north-bound interface (NBI).

\begin{figure}[h!]
\centering
\includegraphics[width=0.7\textwidth]{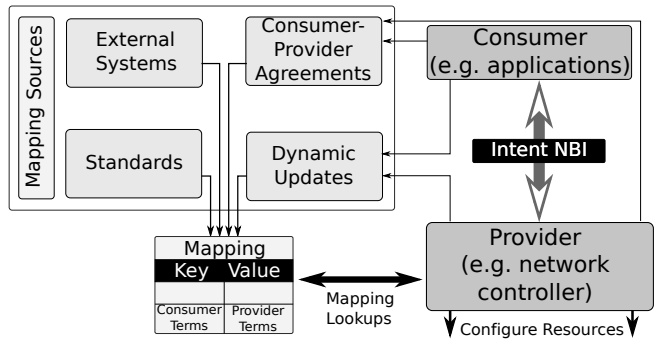}
\caption{\small Intent NBI high-level architecture illustrating the mechanism of frame-of-reference term mapping for intents.}
\label{fig:IntentNBI}
\end{figure}

The \textit{north-bound interface intent framework} (Intent NBI)~\cite{nbi:2016} proposed by the Open Networking Foundation and implemented in the ONOS intent framework~\cite{onos:2016}, adopts this approach.
Intent NBI aims to separate consumer and provider system implementations and simplify consumer-originated requests to provider systems.
This is realized through non-prescriptive and composable requests, independent of network operator implementations and internal policies~\cite{nbi:2016}.
The non-prescriptive property allows the network controller the greatest degrees of freedom in fulfilling service requests and thus facilitates 
conflict resolution, since intent requests do not specify which resources providers must allocate to specific services.
Implementation independence is supported by a mapping mechanism, to provide the bridge between application and controller frame-of-reference terms.
Figure~\ref{fig:IntentNBI} presents a high-level view of the intent NBI approach to translating the policies -- expressed in consumer application terms -- to configurations -- expressed in network controller terms.

The intent NBI currently lacks a mechanism to encapsulate the details of access control over network resources, in order to shield them from installed network applications.
We extend the Intent NBI with a north-bound access control API ($\mathsf{NACA}$) that exposes such higher-level abstractions to applications and implements them on the control plane.
$\mathsf{NACA}$ allows to both produce policy-defined network slices and complement the intent framework with additional attributes, such as physical resource visibility, execution environment access, concurrency, etc.
It is flexible enough to accommodate the various access models of the taxonomy introduced in~\S\ref{subsec:ACS}.

We introduce $\mathsf{NACA}$ by first describing its concepts and underlying mechanisms.
We next delve into the underlying implementation by describing:
the north-bound access control API, which extends the Intent NBI introduced above and allows applications to declare the network resources and type of access necessary for their functionality;
a mechanism to reliably tag application requests with their respective access mask; 
finally, a reference monitor, which ensures that illegal queries issued by applications are detected and discarded even in the event of a network controller compromise.

\subsection{Scalable Access Control for SDN Resources}
\label{subsec:SACS}

Beyond managing access to network resources, the centralized NIB introduced in the SDN model allows to limit access to resources using \textit{partial views} of the system according to various dimensions, such as
geographical or logical placement of resources, visibility of the underlying execution platform, etc.

This approach differs from both network \textit{slicing} and from \textit{network virtualization}.
In~\cite{sherwood:2009,sherwood:2010}, FlowVisor defines slices along any combination of ten packet header fields, including physical, link, network, and transport layers; 
furthermore, such slices can be defined with negation (“all packets but TCP packets with dst port 80”), unions (“ethertype is ARP or IP dst address is 255.255.255.255”), or intersections (“netblock 192.168/16 and IP protocol is TCP”). 
Network virtualization decouples virtual topologies from the physical infrastructure, without exposing the mappings; 
instead, tenants only see their virtual networks~\cite{drutskoy:2013}.
However, neither slicing nor network virtualization can support the rich variety of access models introduced in~\S\ref{subsec:ACS}.

We propose dividing the flowspace, on the north-bound API level, according to: (1) the resource access requests of the subjects and (2) \textit{access masks}, defined -- per subject -- by the MANO component.
Subjects (i.e. the installed applications) declare the resource access requests in a deployment manifest~(\S\ref{subsubsec:NACA}).
Access masks describe limitations to resource access on a higher abstraction level and can depend on the attributes of the network resources themselves, of the environment where they execute, or on the state of the SDN deployment~(\S\ref{subsubsec:tagging}).
Finally, limitations described by the access masks are enforced by $\mathsf{NACA}$ on the controller platform~(\S\ref{subsubsec:refmon}).

This approach -- first outlined in~\cite{paladi:2015} -- is based on a combination of earlier introduced access control approaches, such as capability-based access (CBA)~\cite{fabry:1974,levy:1984}, attribute-based access control (ABAC)~\cite{hu:2013} and policy-based access control (PBAC)~\cite{han:2012}.
CBA is a subject-oriented approach, where a \textit{capability} represents an unforgeable token used to access a resource.
Subjects store their capabilities as sets of pairs (\textit{$x_i$}, $\{R\}$), where \textit{x} represents a resource and $\{R\}$ represents the set of access rights to the resource granted to the subject.
PBAC allows flexible management of access rules using policies -- expressed as sets of rules combined to decide authorization and determine authorization level -- and can be seen as a standardization of ABAC for governance-oriented structures~\cite{bertrand:2015}.
Restricting user access to resources using an access control API has been introduced earlier~\cite{brown:2004,ferguson:2012}, as well as limiting access to data based on higher-level attributes (e.g. based on geographical location~\cite{piccionelli:2010}). 
However, to the best of our knowledge \textit{there is currently no support for access control based on higher-level attributes for SDN controllers}.

We describe the framework allowing attributes to be used as input for access masks that limit access to network resources. 
We \textit{do not} aim to provide an exhaustive list of resource attributes that can serve as input for access masks, since attributes are resource- and implementation-specific.
The rationale behind the proposed approach is based on the portability, performance, elasticity and (multiple) security requirements formulated for VNFs~\cite{etsi:2013}.
We start by defining a resource attribute in the context of $\mathsf{NACA}$:

A \textit{resource attribute} is a property of the SDN resource that can be used to describe access constraints on the respective resource.
The values of a collection of resource attributes can determine the type and scope of access to an SDN resource granted to a subject.

Table~\ref{tab:resource-mask-attributes} outlines several examples of attributes (beyond trivial ones such as e.g. instance name or identifier).
Recall the example introduced by Kephart~\cite{kephart:2004} and described above.
The network resource attributes presented in Table~\ref{tab:resource-mask-attributes} can be used to create access restrictions on the goal policy level, without specifying details about either the deployment itself or the various protocols that it uses for internal communication among the components.

\begin{table}[h]
 \scriptsize 
\begin{center}
\caption{Example attributes and clarifications}
\label{tab:resource-mask-attributes}
  \begin{tabulary}{\textwidth}{L L}

		Placement							& Geogr./logical placement of accessible resources	  		  \\
  	\hline
		Scope     							& Aggregate vs domain-specific access 			  			  \\
  	\hline
		Time  								& Continuous updates vs discrete updates			  			  \\
  	\hline
		Jurisdiction							& Jurisdictional placement of accessible resources 			  \\
  	\hline
		Physical resource visibility 		& Visibility of underlying execution environment 	  			  \\		 		
  	\hline
	    Execution environment access 		& Direct vs mediated access to physical resources	  			  \\	
  	\hline
		Resource modification types 			& Read state vs modify state						  			  \\
  	\hline
		Concurrency							& Exclusive (locking) or non-exclusive access      			  \\
  	\hline
	    Authority delegation                 & Ability to delegate access capabilities          			  \\ 	 			
  \end{tabulary}  
\end{center}
\end{table}

In a typical workflow, network operators define for each available resource $R$ (and based on its attributes) a set of \textit{resource access rules}~(\ref{eq:resourceAccessRules}).
\begin{equation} \label{eq:resourceAccessRules}
R_i = \{rar_i^1, ...,\ rar_i^n\};\ R_j = \{rar_j^1,...,\ rar_j^n\}; ...; \ R_m = \{rar_m^1,...,\ rar_m^n\}
\end{equation}
The resource access rules contain values for the relevant attributes (e.g. as in Table~\ref{tab:resource-mask-attributes}) of each resource.

A candidate application declares through an \textit{application deployment manifest} the network resources it requires for its functionality.
Besides a structured enumeration of the required resources, the manifest optionally contains the types of actions to be performed.
The set of requested resources must be a subset of available resources advertised by the network controller~(\ref{eq:ResourceDefinition}).
\begin{equation} \label{eq:ResourceDefinition}
\{r_i, r_j, ..., r_m\} \in AvailableResources
\end{equation}
The resource enumeration in the deployment manifest (DM) consists of a list of tuples, where each tuple contains a requested resource and a set of actions on the resource that the application requires for its functionality.
This has the form $\langle resource, actions \rangle$, as shown in~(\ref{eq:RequestDefinition}):
\begin{equation} \label{eq:DeploymentManifest}
DM = \{\langle r_i, \{a_1, a_2, ..., a_n\} \rangle, \langle r_j, \{a_1, a_2, ..., a_n\}\rangle,...,\langle r_m, \{a_1, a_2, ..., a_n\}\rangle\}
\end{equation}
The actions listed in the deployment manifest (\ref{eq:DeploymentManifest}) are resource and implementation specific:
Ferguson et al. describe two types of actions -- \textit{read} and \textit{write}~\cite{ferguson:2013};
Klaedtke et al. propose an expanded action set, including permissions for reading statistics (\texttt{stat}), requesting information about an object (\texttt{config\_read}), modifying the state of an object (\texttt{config\_mod}), as well as for subscription permissions (\texttt{subscr})~\cite{klaedtke:2014}.
Investigation of a comprehensive taxonomy of \textit{actions} applicable to SDN resources is out of the scope of this paper and left for future work.

The MANO component computes the access mask.
First, it selects the resource access rules applicable to the requested resources and relevant to the application and builds an operator policy set~(\ref{eq:RequestDefinition}):
\begin{equation} \label{eq:RequestDefinition}
OP = \{\langle rar_i^1, ..., rar_i^n \rangle, \langle rar_j^1, ..., rar_j^n \rangle, ... , \langle rar_m^1, ..., rar_m^n \rangle\}
\end{equation}
Next, it applies a function $f_{AM}$ to map each element in the operator policy set $OP$ to at most one element in DM~(\ref{eq:function}).
Note that depending on the number and scope of the resource access rules, $f_{AM}$ is either a surjective, bijective or injective mapping of resource access rules to requested resources.
\begin{equation} \label{eq:function}
f_{AM}: OP \longrightarrow DM
\end{equation}
The resulting access mask $AM$ is a three element tuple of the form $\langle resource,$ $actions,$ $resourceAccessRules\rangle$ (\ref{eq:AccessMask}).
Note that there is no requirement that resource access rules describe all possible attributes of a resource.
\begin{equation} \label{eq:AccessMask}
AM = \{\langle r_i, \{a_1,...,a_n\}, \{rar_i^1,...,rar_i^n\} \rangle,..., \langle r_m, \{a_1...a_n\}, \{rar_m^1, ..., rar_m^n\}\rangle \}
\end{equation}
The MANO component maintains a dictionary of installed applications and their resource masks.
For each new installed application, conflict detection is done by recursively checking matching resources, matching resource access rules and finally matching resource attribute values.
Applications with conflicting resource requests can only be installed once the conflict is resolved.
This approach to specifying access masks allows to implement any of the resource access models described in the taxonomy introduced in~\S\ref{subsec:ACS}.

The resulting access mask is communicated to the \textit{\textit{Request Tagger}}, which -- in combination with the \textit{Reference Monitor} -- ensures that the respective application cannot access resources outside the access mask, as described below.

\subsection{$\mathsf{NACA}$ Internals}
\label{subsec:internals}

\begin{figure}[t]
\centering
\includegraphics[width=0.8\textwidth]{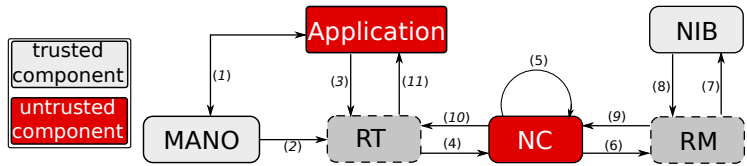}
\caption{\small $\mathsf{NACA}$ communication model.
\textit{Components:} MANO: management and orchestration component;
RT: request tagger; NC: network controller; RM: reference monitor; NIB: network information base.
\textit{Communication steps:} 
(1) MANO deploys application and computes the access mask; 
(2) Send access masks of installed applications to RT; 
(3) submit application request;
(4) compute the tag and an alias, communicate them to RM;
(5) forward the request, along with the access mask and alias, to the NC;
(6) compile request to set of queries; 
(7) forward compiled queries reference monitor;
if verification successful (8), forward to NIB (9); return result (10-13).}
\label{fig:NAAC-communication}
\end{figure}

To implement support for capability-based access control policies in $\mathsf{NACA}$, we have applied the Policy Core Information Model (PCIM) IETF specification~\cite{moore:2001}, which describes an object-oriented information model for representing policy information.
The architecture of $\mathsf{NACA}$ is aligned with the PCIM approach.
The flow of intents issued by installed applications is illustrated in Figure~\ref{fig:NAAC-communication}.
The depicted components can be directly mapped to the elements of the SDN architectural model (see Figure~\ref{fig:SDN_architecture}), with the exception of \textit{Request Tagger} (RT) and Reference Monitor (RM), both introduced as supporting mechanisms for $\mathsf{NACA}$. 
Figure~\ref{fig:NAAC-communication} also depicts the trust relationship between the included components, according to the threat model in~\S\ref{subsec:adv_model}: 
the MANO component, \textit{Request Tagger}, \textit{Reference Monitor} and the NIB are considered to be trusted, while the application(s) and network controller are untrusted.

In a vanilla approach, application intents are forwarded to the network controller which maps them to its internal frame of reference (recall Figure~\ref{fig:IntentNBI}), compiles them to a set of configuration instructions and applies the instructions on the network information base.
We next motivate the introduction of additional components and describe their functionality.

\subsubsection{North-bound Access Control API}
\label{subsubsec:NACA}

Network applications vary significantly in their intended functionality.
While software development best-practices emphasize data and function encapsulation~\cite{greenfield:2003}, application developers cannot be expected to either 
develop applications that efficiently use limited SDN resources or have an understanding of resource partitioning within any particular SDN deployment.
On the other hand, it is reasonable to expect that application developers are interested in explicitly specifying the complete set of \textit{sufficient} resources for correct application functionality.
This may include elements of all resource categories introduced in~\S\ref{subsec:sys_model} -- device, control, and data resources.

A network operator uses the deployment manifest (among other parameters) to decide whether a certain application should be installed or included in the application service catalog.
Listing~\ref{lst:depman} shows an example deployment manifest in a notation based on the standardized OASIS eXtensible Access Control Markup Language (XACML) Version 3.0~\cite{oasis:2013} (pruned for clarity and brevity).

\noindent\begin{minipage}[t]{.45\textwidth}
\lstset{basicstyle=\tiny,style=smallxml}
\lstset{language=XML, morekeywords={encoding,AttributeValue,Actions,Action,ResourceAttributeDesignator, Resource, Resources,ResourceMatch, AnyOf, AllOf, Match,AttributeDesignator}}
\begin{lstlisting}[caption=Deployment manifest fragment,label=lst:depman,frame=tlrb]
<AnyOf>
 <AllOf>
  <Match MatchId="string-equal">
	<AttributeValue>dataplane topology</AttributeValue>
     <AttributeDesignator AttributeId = "resource-id" Category="resource"/>
  </Match>
 </AllOf>
</AnyOf>
<AnyOf>
<AllOf>
  <Match MatchId="string-equal">
    <AttributeValue>read</AttributeValue>
     <AttributeDesignator AttributeId = "action-id" Category="action"/>
  </Match>
</AllOf>
<AllOf>
  <Match MatchId="string-equal">
    <AttributeValue>modify</AttributeValue>
    <AttributeDesignator AttributeId = "action-id" Category="action"/>
  </Match>
</AllOf>
</AnyOf>
\end{lstlisting}
\end{minipage}
\begin{minipage}[t]{.53\textwidth}
\lstset{basicstyle=\tiny,style=smallxml}
\lstset{language=XML, morekeywords={encoding,AttributeValue,Actions,Action,ResourceAttributeDesignator, Resource, Resources,ResourceMatch, AnyOf, AllOf, Match,AttributeDesignator}}
\begin{lstlisting}[caption=Resource mask fragment based on deployment manifest, label=lst:accmas,frame=tlrb]
<AnyOf>
 <AllOf>
   <Match MatchId="string-equal">
	<AttributeValue> dataplane topology </AttributeValue>
      <AttributeDesignator AttributeId="resource-id" Category="resource"></AttributeDesignator>
   </Match>
   <Match MatchId="string-equal">
	<AttributeValue DataType="string">region-A</AttributeValue>
      <AttributeDesignator AttributeId="jurisdiction" Category="resource"></AttributeDesignator>
   </Match>
   <Match MatchId="string-equal">
	<AttributeValue DataType="string">dataplane topology</AttributeValue>
      <AttributeDesignator AttributeId="resource-id" Category="resource"></AttributeDesignator>
   </Match>
 </AllOf>
</AnyOf>
<AnyOf>
 <AllOf>
   <Match MatchId="string-equal">
     <AttributeValue DataType="string">read</AttributeValue>
                 <AttributeDesignator AttributeId="action-id" Category="action">
     </AttributeDesignator>
   </Match>
 </AllOf>
</AnyOf>
\end{lstlisting}
\end{minipage}

Once installed, an application containing in its deployment manifest the fragment from Listing~\ref{lst:depman} can read and modify the topology on the data plane.
While the application requires read permissions to the network topology, revealing the \textit{entire} topology might be unacceptable to the network operator.
Instead, the operator may consider allowing the application to access only a \textit{restricted subset} of the topology and present the rest as a black box.
Therefore, the operator uses the MANO component to apply a simple resource access rule to the resource ``\texttt{dataplane-topology}''.
The rule comprises only one attribute -- jurisdiction -- with the value ``\texttt{region-A}''.
Once applied, the resource access rule reduces the visibility of the topology for the application exclusively to the selected region, while the rest of the topology is not reported.

To represent access masks, we extend the OASIS XACML notation~\cite{oasis:2013} (see Listing~\ref{lst:accmas}).
The resource mask example in Listing~\ref{lst:accmas} allows an application to access the topology of the SDN deployment;
however, it is only limited to the resources located in region ``A''.
The MANO component communicates the access masks for each installed application to the \textit{Request Tagger}.

\subsubsection{Request tagging}
\label{subsubsec:tagging}
Application requests must be \textit{authenticated} and \textit{tagged} with an access mask prior to reaching the network controller.
While such functionality can be implemented by the network controller, we have chosen a modular approach for the following reasons:
first, authentication and tagging are generic functions that can be implemented independently from a deployment-specific network controller;
second, this approach allows to minimize the network controller modifications, required to implement $\mathsf{NACA}$;
finally, the network controller is a high-value, high-risk target which may contain API vulnerabilities which can be exploited by untrusted applications;
moreover, query parsing is a common attack vector~\cite{spath:2016} which could be used to corrupt the network controller.
The reasons above motivate the introduction of a \textit{\textit{Request Tagger}} -- a pre-processing component implementing access control on the north-bound interface, resilient to a potentially compromised network controller (Figure~\ref{fig:NAAC-communication}).

\begin{figure*}[]
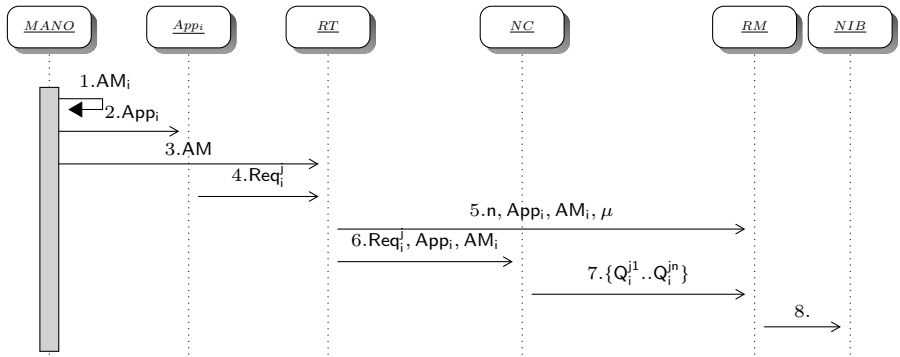

\begin{center}
\tikzstyle{every node}=[font=\scriptsize]
	\begin{sequencediagram}
		\tikzset{inststyle/.append style={
        drop shadow={top color=gray,bottom color=white}, 
        scale=0.69,
        rounded corners=1.0ex},
        scale=0.8,
      			}
		\newthread{mano}{$MANO$} 
     	\newinst[1]{app}{$App_i$} 	 
     	\newinst[1]{rt}{$RT$} 	 
     	\newinst[2]{nc}{$NC$} 	 
     	\newinst[2.7]{rm}{$RM$} 	 
     	\newinst[0.2]{nib}{$NIB$} 	 
   				\callself{mano}{$1.\mathsf{AM_i}$}{}
				\mess{mano}{$\hspace{10pt} 2. \mathsf{App_i}$}{app}
				\mess{mano}{$3. \mathsf{AM}$}{rt}
				\mess{app}{$4. \mathsf{Req_i^j}$}{rt}
				\mess{rt}{$5. \mathsf{n, App_i, AM_i, \mu}$}{rm}				
				\mess{rt}{$6. \mathsf{Req_i^j, App_i, AM_i}$}{nc}
				\mess{nc}{$7.\mathsf{\{Q_i^{j1}..Q_i^{jn}\}}$}{rm}			
				\mess{rm}{$8.$}{nib}
    \end{sequencediagram}
\caption{\small Message flow in $\mathsf{NACA}$ (we assume communication over a secure channel).
\textit{Components}: MANO: management and orchestration component;
RT: request tagger; NC: network controller; RM: reference monitor; NIB: network information base.
\textit{Flow}:
(1)~MANO computes access mask and (2)~deploys the app; 
(3)~MANO distributes to \textit{Request Tagger} the access mask for $\mathsf{App_i}$;
once the application $\mathsf{App_i}$ issues a request~(4), RT computes a tag and communicates it to the RM~(5), prior to forwarding the request to the NC where the request is processed and compiled into a set of queries~(6), forwarded to the RM~(7).
RM verifies the queries, computes a MAC over each query with a nonce -- producing $\mathsf{\{\{Q_i^{j1}, u^1, \mu_i^{j1}\}..\{Q_i^{jn}, u^n, \mu_i^{jn}\}\}}$ and forwards them to the NIB~(8).}
\label{fig:naca-intra}
\end{center}
\end{figure*}

At deployment time, the MANO component communicates to the \textit{Request Tagger} a set of tuples describing the installed applications and their access mask;
the set of tuples is updated for every new installed application.
We consider the \textit{Request Tagger} a trusted component (see discussion in~\S\ref{subsec:compis}) and assume the integrity of such messages can be ensured and reliably verified.
Application requests on the north-bound interface must contain a reliably verified application identifier, the set of requested resources (i.e. elements included in the application deployment manifest), and an intent that can be compiled into one or more implementable queries to the NIB.
An incoming request is pre-processed by the \textit{Request Tagger}, which:
\begin{enumerate}
	\item verifies the authenticity of the request;
	\item matches the application instance -- identified by ($\mathsf{App_{i}}$) -- with one of the access masks ($\mathsf{AM_i}$) earlier communicated by the MANO component;
	\item assigns a unique identifier to the request ($\mathsf{Req_{i}^j}$);
	\item tags the request with the identified access mask.
\end{enumerate}
For step 4, the \textit{Request Tagger} constructs a tag~(\ref{eq:tag}) which is a MAC over the following elements:
identifier of the requesting application ($\mathsf{App_{i}}$);
unique identity of the request ($\mathsf{Req_{i}^j}$);
application access mask $\mathsf{AM_i}$;
monotonically increasing sequence counter $\mathsf{n}$.
The MAC value is computed using a shared key $\mathsf{K}$ distributed by the MANO component at deployment time to the \textit{Request Tagger} and \textit{Reference Monitor} (see Figure~\ref{fig:naca-intra}).
\begin{equation} \label{eq:tag}
\mathsf{\mu\ =\ MAC(K, (App_{i}, Req^j_{i}, AM_{i}, n) )}
\end{equation}

Next, the application request is forwarded to the network controller without further processing, along with $\mathsf{AM_{i}}$ -- the application access mask.
The network controller processes the request, applies operator-defined policies and compiles a request $\mathsf{Req^j_{i}}$ into a set of discrete queries $\{Q_i^{j1}..Q_i^{jn}\}$.
Note that the compiled queries contain the identifier of the requesting application ($\mathsf{App_{i}}$) and the unique request identifier ($\mathsf{Req_{i}^j}$).
Internals of network controller processing are out of the scope of this description and can be found in~\cite{berde:2014,onos:2016}.

\subsubsection{Access Reference Monitor}
\label{subsubsec:refmon}
The \textit{queries} produced by the network controller describe discrete changes made to e.g. the configuration of network elements or flow tables on the forwarding plane.
However, a network controller containing API vulnerabilities, implementation bugs or maliciously modified through an attack~\cite{dhawan:2015} may produce queries that invoke resources outside of the application's access mask.
To address this, we introduce a \textit{Reference Monitor} as a discrete component (similar to the approach described in~\S\ref{subsubsec:tagging}).
It validates the network controller output (i.e. queries ready for execution on the NIB) and ensures that illegal application queries fail to reach the NIB.
{\textit{Illegal application queries}} are those that on behalf of an application invoke resources outside of its access mask.

The workflow of the \textit{Reference Monitor} is as follows.
At deployment time, the MANO component pre-seeds the \textit{Reference Monitor} with the key material (e.g. public key infrastructure certificate) required to establish an authenticated, confidentiality and integrity-protected communication channel, as well as a shared key $\mathsf{K}$ used for authenticating incoming request tags.
This communication channel is maintained throughout the component lifetime.
In-transit component protection can be ensured using mechanisms described in~\cite{paladi:2016b}.

\paragraph{Request matching}
For each application request, the \textit{Request Tagger} communicates to the \textit{Reference Monitor},  over a secure channel, the tag $\mu$ computed according to (\ref{eq:tag}), along with the application identifier ($\mathsf{App_i}$), its current access mask ($\mathsf{AM_i}$) and a request counter $\mathsf{n}$.
Similarly, the network controller forwards to the \textit{Reference Monitor} for verification all queries, produced from application requests according to the respective access mask.

Upon receiving a set of queries from the network controller, the \textit{Reference Monitor} first verifies the freshness of the monotonically increasing counter $\mathsf{n}$ (it is expected that a correctly functioning network controller compiles requests in first-in-first-out order).
It next computes a tag $\mu'$ according to operation~(\ref{eq:tag}) and compares the result with the received tag $\mu$;
execution only continues if $\mu'=\mu$.
Finally, it parses the queries and verifies them against the access mask of the application.
This is done  by checking that:
\begin{itemize}
	\item queries exclusively invoke resources enumerated in the access mask;
	\item resources are invoked according to the actions specified in the access mask.
\end{itemize} 

\paragraph{Query invalidation}
While duration of request compilation into queries can vary depending on the complexity of the requests, a malicious network controller may delay or reorder output of queries.
To prevent the reuse of a more permissive access mask for lower-privileged applications, we use a \textit{sliding window} for invalidating the computed queries. 
The approach is as follows:
a delayed request $\mathsf{Req_a^1}$ issued by application $\mathsf{App_a}$ can be followed by a limited number of requests $\mathsf{Req_a^2...Req_a^n}$ issued by the same application, while a delay of $\mathsf{Req_a^{n+1}}$ requests from the same application invalidates the entire batch (the limit is configuration-specific);
delay of a request $\mathsf{Req_a^1}$ issued by $\mathsf{App_a}$ followed by a request $\mathsf{Req_b^1}$ issued by $\mathsf{App_b}$, as well as reordering of the requests issued by $\mathsf{App_a}$ and $\mathsf{App_b}$, invalidates both requests.

The verification by the \textit{Reference Monitor} adds an essential element of the \textit{direct explicit assignment} resource access model introduced in~\S\ref{subsec:ACS}:
access is only permitted to SDN resources that have been explicitly described and over actions that have been explicitly listed.
We exemplify this below.

Recall the access mask fragment in Listing~\ref{lst:accmas}.
Example~\ref{ex:illegalQ} shows a query to install a new flow between source IP $\mathsf{w}$, port $\mathsf{w1}$ and destination IP $\mathsf{x}$, port $\mathsf{x1}$ over the user datagram protocol.
The \textit{Reference Monitor} would trivially reject this query since it invokes a resource (flow) that is \textit{not} enumerated in the deployment manifest (and hence would \textit{not} be present in the access mask).
\begin{exmp}{\textit{New flow installation query}}\\   \label{ex:illegalQ}
$\langle \mathsf{flow,\ allow,\ srcIP=w, dstIP=x}$ $\mathsf{proto=UDP, srcPort=w1, dstPort=x1}\rangle$
\end{exmp}

Depending on the size and complexity of the query set, verifying queries may require multiple interactions between the \textit{Reference Monitor} and NIB to learn the attributes of the invoked resources.
In Example~\ref{ex:legalQ} the application requests a list of logical termination points that bound a node `$\mathsf{j}$':
\begin{exmp}{\textit{Application request}}\\ \label{ex:legalQ}
$\langle \mathsf{topo,\ node_j,\ ltpRefList}\rangle$
\end{exmp}
To verify this query, the \textit{Reference Monitor} must first obtain the topology reference of the respective node ($\mathsf{node_j}$), identify its jurisdiction and compare it with the jurisdiction in the access mask (\texttt{region-A} in Listing~\ref{lst:accmas}).
If $\mathsf{node_j}$ is located in \texttt{region-A}, a list of logical termination points (also located in the same jurisdiction) is returned to the caller.
Otherwise the request is denied.

In case of an access mask mismatch -- which may indicate an intent compiler bug or a compromise of the network controller -- the \textit{Reference Monitor} drops the illegal query and notifies the MANO component to take an appropriate mitigation action.
While the access mask verification prevents illegal queries, a compromised network controller may attempt other attacks as well, such as attempt to circumvent the \textit{Reference Monitor} by compiling a request into a set of unintended queries (which nevertheless comply with the access mask).
Considering that a vulnerability in the network controller must be exploited (or triggered) through a request from a malicious application, such an attack is prevented by the sliding window invalidating delayed or reordered queries.
This prevents the adversary from inserting malicious queries into the set of queries compiled from the request of a benign application.
The ordering, origin and privileges of submitted requests -- communicated from the \textit{Request Tagger} to the \textit{Request Manager} -- provide the information necessary for invalidating queries in case of suspected network controller compromise.
Note however that these steps help prevent exploitation of a compromise and do not aim to detect all occurrences of compilation errors.
Likewise, the steps above do not prevent malicious actions by code built into the intent compiler.
This problem (first formulated in~\cite{thompson:1984}) is out of the current scope.

For valid requests, the \textit{Reference Monitor} computes a MAC over each query and a nonce `$\mathsf{u}$' using the key $\mathsf{K_{NIB}}$, shared between \textit{Reference Monitor} and NIB (\ref{eq:mac-nib}). 
\begin{equation} \label{eq:mac-nib}
\{\mu_i^{j1}..\mu_i^{jn}\} = \mathsf{MAC(K_{NIB},({Q_i^{j1}, u^1}))}..\mathsf{MAC(K_{NIB},({Q_i^{jn}, u^n}))}
\end{equation}
Finally, the \textit{Reference Monitor} forwards the queries, nonces and computed MACs to the NIB, which first recomputes the MACs and only processes queries with verified integrity and a fresh nonce.

\section{Implementation}
\label{sec:implementation_onos}

We have implemented $\mathsf{NACA}$ as an extension to ONOS, a popular open-source SDN controller~\cite{berde:2014}.
ONOS includes an \textit{intent} framework, allowing applications to specify their \textit{network control desires} as policies rather than specific mechanisms~\cite{onos:2016}.
This raises the abstraction level for applications from specifying details about \textit{how} the SDN infrastructure configuration should be updated to achieve certain functional changes (e.g. by updating network flows, instantiating new applications, etc.) to conveying through high-level intents \textit{what} functionality should be enabled.
This shift is equivalent to replacing the need to specify network mechanisms acting through OpenFlow operations with tools that are more durable and robust in the face of topology changes.
Once created by an application, \textit{intents} become immutable objects communicated to the ONOS core which -- when installed -- alter the network state.

\subsection{Extending the ONOS Intent Framework}
\label{subsec:exoif}

\begin{figure}[t]
\centering
\includegraphics[width=0.6\textwidth]{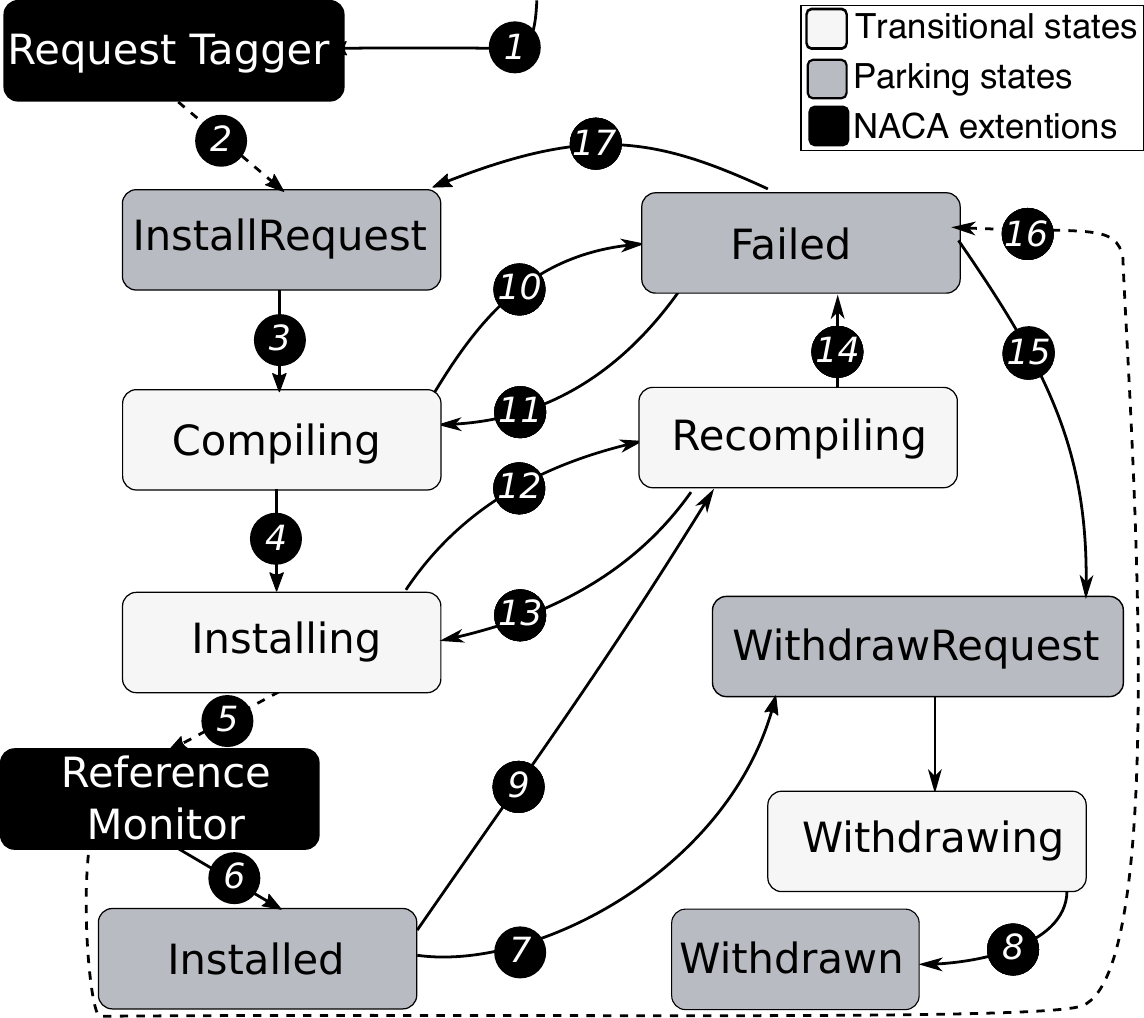}
\caption{\small $\mathsf{NACA}$ Extended intent framework state diagram.
\textit{Transitions:} (1) application request; (2) tagged query request; (3) submit for compilation; (4) compile succeeded; (5) install succeeded; (6) verify access compliance; (7) withdrawal installed request; (8) remove topo or flow event; (9) add/update topo event; (10) compile failed; (11) retry compile; (12) install failed; (13) retry install; (14) compile failed or same result; (15) withdrawal of failed requests; (16) reference monitor rejected request; (17) retry install intent.}
\label{fig:intent-states}
\end{figure}

In ONOS, intent objects contain the identifier of the issuing application \newline
(\texttt{ApplicationId}) and are uniquely identifiable through the intent identifier (\texttt{IntentId}), generated when the query object is created in the intent framework based on an application request.
An ONOS intent is additionally described by the following elements:
the required device resources (a collection of \texttt{networkResource} objects, similar to the the $\mathsf{NACA}$ queries described above);
the intent \texttt{priority} and \texttt{constraints} prescribed by the application (criteria enumerating e.g. packet header fields or patterns describing slices of traffic);
instructions describing actions to be applied to a slice of traffic.
To implement $\mathsf{NACA}$, we extend the intent framework with two components -- \texttt{RequestTagger} and \texttt{ReferenceMonitor}.
Figure~\ref{fig:intent-states} depicts the state transition diagram for compilation of application intents~\cite{onos:2016}, along with the $\mathsf{NACA}$ extensions.

The \texttt{RequestTagger} pre-processes intents submitted by applications.
It first reliably determines the identity of the issuing application by checking the intent signature and certificate, and assigns a corresponding \texttt{ApplicationId}.
It next identifies the access mask matching the \texttt{ApplicationId}, computes the tag $\mathsf{T}$ (see~\ref{eq:tag-t}), communicates it to the \texttt{Reference Monitor} and transitions to the \texttt{InstallRequest} state (transition \circled{\color{white}2} in Figure~\ref{fig:intent-states}).
\begin{equation} \label{eq:tag-t}
\mathsf{T\ =\ MAC(K, (ApplicationId,\ IntentId} \mathsf{networkResource,\ priority,\ AM,\ n) )}
\end{equation}

In ONOS, we implemented the \texttt{RequestTagger} as a state of the intent framework, invoked directly from the \texttt{IntentManager} class.
The intent framework compiles requests issued by the applications into queries according to constraints specified in the access mask.
Prior to executing the compiled query on the NIB and transitioning to the \texttt{Installed} state, the execution flow transitions to the \texttt{ReferenceMonitor} (transition \circled{\color{white}5} in Figure~\ref{fig:intent-states}).
The \texttt{ReferenceMonitor} first verifies tag $\mathsf{T}$ by recomputing the $\mathsf{MAC}$:
\begin{equation} \label{eq:rec-mac}
\mathsf{MAC(K, (ApplicationId,\ IntentId} \mathsf{networkResource,\ priority,\ AM,\ n) )}
\end{equation}
It next verifies that the queries fall into the validation sliding window, by checking that all earlier expected  queries have been output (sliding window for delayed requests was configured to \texttt{n=2}) and that queries have the expected source application and access mask.
Finally, it checks that the compiled queries do not violate the access mask assigned by the MANO component, as described in~\S\ref{subsubsec:refmon}: the \texttt{ReferenceMonitor} parses the query to check if all invoked resources are included in the access mask, if queries follow the limits of the resource attributes, as well as whether the queries invoke only actions allowed by the access mask.
A MAC is computed over valid queries and submitted -- along with the requests -- to the NIB.

Similar to the \texttt{RequestTagger} above, the \texttt{ReferenceMonitor} is implemented as a state of the ONOS intent framework.
Every intent must transition to the \texttt{ReferenceMonitor} state before the resulting compiled queries can be applied.
Queries submitted directly to the NIB are ignored, as they lack a valid MAC that must be computed by the \texttt{ReferenceMonitor}.

The \texttt{Constraint} implementation in ONOS allows applications to formulate filters which are applied on the submitted intents.
We leverage this implementation to apply the access masks assigned by the MANO component -- the \texttt{ReferenceMonitor} ensures that the candidate intents comply with the access mask constraints.
If an access mask violation is detected, the \texttt{ReferenceMonitor} transitions to the \texttt{Failed} state (transition \circled{\color{white}16} in Figure~\ref{fig:intent-states}).

\subsection{Component Isolation}
\label{subsec:compis}
\textit{Request Tagger} and \textit{Reference Monitor} are of central importance in verifying access of untrusted applications to limited, potentially confidentiality and integrity sensitive resources.
To protect them from a malicious network controller, the \textit{Request Tagger} and \textit{Reference Monitor} are executed in a trusted execution environment \textit{isolated} from a potentially malicious underlying operating system~\cite{schuster:2015,paladi:2016b,arnautov:2016}.
We adopt the definition of trust from~\cite{etsi:2014}, namely ``confidence in the integrity of an entity for reliance on that entity to fulfill specific responsibilities''.
A trusted execution environment can be created using operating system level virtualization~\cite{soltesz:2007}, platform virtualization~\cite{mccune:2010}, or using hardware-assisted isolated execution environments~\cite{zhang:2016} -- such as ARM TrustZone~\cite{arm:2016}, Intel SGX enclaves~\cite{mckeen:2013} or  AMD secure memory encryption~\cite{ledacky:2016}.
The exact implementation approach depends of multiple factors, such as threat model, deployment context, acceptable performance penalty or hardware capabilities.
In the current implementation, as further described in Section~\ref{sec:evaluation}, we have chosen operating system level virtualization to separate the trusted components -- such as \texttt{RequestTagger}, \texttt{ReferenceMonitor} and the NIB -- from the vulnerable network controller.
This approach allows to isolate the process and address spaces of the components, while avoiding excessive overhead.


\section{Evaluation}
\label{sec:evaluation}
We next evaluate the security (\S\ref{subsec:security-evaluation}) and performance (\S\ref{subsec:perf-evaluation}) of $\mathsf{NACA}$.

\subsection{Security Evaluation}
\label{subsec:security-evaluation}
To evaluate the security of $\mathsf{NACA}$, we review the attack vectors available to the adversary, considering the the system model defined in \S\ref{subsec:sys_model} and the threat model defined in~\S\ref{subsec:adv_model}.
In the context of $\mathsf{NACA}$, the following attack vectors available to an adversary:
\begin{itemize}
	\item \textit{Vector 1} - attacks using arbitrary requests of network resources;
	\item \textit{Vector 2} - packet tampering;
	\item \textit{Vector 3} - disrupt or degrade network connectivity;
	\item \textit{Vector 4} - trigger software vulnerabilities with malformed or crafted packets.
\end{itemize}

\paragraph{Vector 1}
The adversary may attempt to submit to the network controller arbitrary resource requests in order to exploit access control vulnerabilities and obtain unauthorized access to network resources. 
This is one of the attack vectors that $\mathsf{NACA}$ aims to protect against.
First, by allowing only \textit{whitelisted} resource requests (filtered using the request mask) and discarding all other requests, the request tagger throttles the number of requests that reach the network controller.
Second, the request mask is defined by the network administrator at the time of enrolling the network application into the deployment; 
this allows to set a restrictive access mask on unverified or untrusted) network applications to prevent abuse.

\paragraph{Vector 2}
The adversary may attempt to tamper with the packets exchanged on the management network between the network elements in the SDN infrastructure.
This class of attacks in SDN infrastructure has been earlier described in~\cite{paladi:2015b}.
Authenticating and encrypting the communication occurring on the management network between the network elements is (as per the assumption in \S\ref{subsec:sys_model}) prevents such attacks.

\paragraph{Vector 3}
The adversary may attempt to cause a Denial-of-Service (DoS) attack on the SDN infrastructure by disrupting or degrading network connectivity.
In the case of an adversary that can drop arbitrary packets on the management network $\mathsf{NACA}$ does not prevent such attacks.
The situation is different in the case when the adversary cannot drop arbitrary packets on the network and launches attacks on the SDN infrastructure through queries issued from network applications.
In this case $\mathsf{NACA}$ limits the capability to launch a DoS attack by restricting the set of available network resource requests with the resource mask. described in \S\ref{subsec:SACS}.

\paragraph{Vector 4}
Given full access to the north-bound API, the adversary is able to send maliciously crafted packets to trigger vulnerabilities in the packet processing code of network elements, as shown in~\cite{thimmaraju:2018}.
As suggested in~\cite{paladi:2017d}, this attack can be contained by running the potentially vulnerable component in an isolated execution environment that ensures code integrity, such as SGX enclaves (see ~\ref{subsec:compis}.
While this does not exclude code vulnerabilities and hence does not prevent the attack per se, isolated execution and code integrity monitoring prevents the adversary from exploiting the attack and gaining control of additional assets. 

Considering the above analysis of the identified attack vectors, an adversary is only left with the ability to conduct DoS attacks on the SDN infrastructure, limited to the situation when the adversary can drop arbitrary packets on the management network.
However, preventing DoS is out of the scope of $\mathsf{NACA}$ and can be addressed though other technical and administrative mechanisms.

\subsection{Performance Evaluation}
\label{subsec:perf-evaluation}

\begin{figure}[]
\centering
\includegraphics[width=0.8\textwidth]{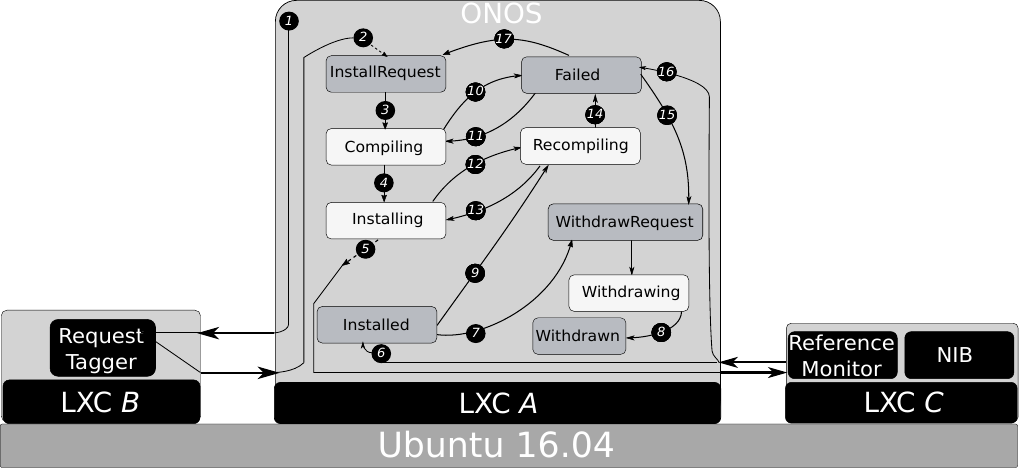}
\caption{\small $\mathsf{NACA}$ testbed}
\label{fig:testbed}
\end{figure}

We evaluated $\mathsf{NACA}$ in a virtualized testbed, as illustrated in Figure~\ref{fig:testbed}.
To limit the influence of system configuration options on the results of the evaluation, we chose a lightweight approach to isolation between the components.
We used Linux Containers~\cite{felter:2015} to isolate the address and memory spaces of the $\mathsf{NACA}$ components.
We deployed the modified ONOS controller, \texttt{RequestTagger} and \texttt{ReferenceMonitor} in three separate containers: LXC~A, LXC~B, and LXC~C respectively.
Note that this deployment choice is not binding and allows for alternative isolation approaches.
The testbed containers were deployed on a Ubuntu 16.04 VirtualBox\footnote{Oracle VirtualBox \url{https://www.virtualbox.org/wiki/VirtualBox}} virtual machine, with 1 CPU and 8 GB memory, default paravirtualization interface.

\begin{figure}[h!]
\centering
  \centering
	\includegraphics[width=0.7\textwidth]{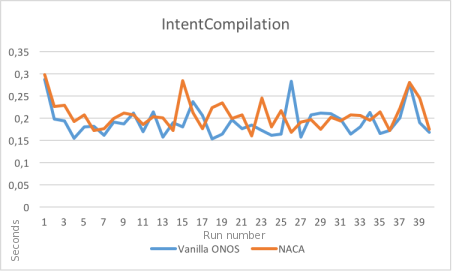}
	\caption{\texttt{compileIntent}, time in s.}
	\label{fig:compileIntent}
\end{figure}

To support backward compatibility and with reproducible results, we evaluated $\mathsf{NACA}$ using the intents and test case coverage available in ONOS.
We chose three parameters to test: intent compilation, intent submission and combined intent compilation and submission.
Figure~\ref{fig:compileIntent} illustrates the performance of intent compilation over 40 test runs.
Here the $\mathsf{NACA}$ extensions induce a 9\% increase on the median intent compilation time (8\% mean increase).
The overhead is caused by tagging the query request (see transition 1 - 2 in figure~\ref{fig:intent-states}).
We believe this overhead can be reduced if the \textit{Request Tagger} is deployed in an isolated execution environment (e.g. SGX enclaves~\cite{mckeen:2013}) rather than a lightweight virtualization container.
We leave this optimization to future work and improved prototype implementations.

\begin{figure}[h]
\centering
  \centering
	\includegraphics[width=0.7\textwidth]{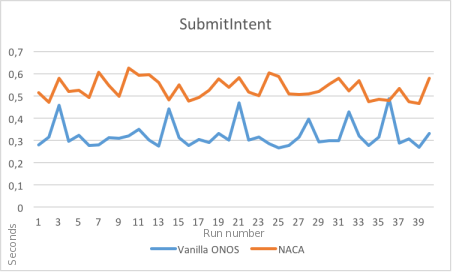}
	\caption{\texttt{submitIntent}, time in s.}
	\label{fig:submitIntent}
\end{figure}

We next evaluated the performance of the entire intent installation chain.
Figure~\ref{fig:submitIntent} illustrates the performance of intent installation over 40 test runs.
Intent installation includes the entire flow from request submission by the application to the queries accepted by the NIB and includes the intent compilation steps evaluated above.
In this case, the mean performance overhead is up to 66\% and reflects implementation and deployment decisions (including choice of isolation mechanism and target deployment platform).

\begin{figure}[h]
\centering
  \centering
	\includegraphics[width=0.7\textwidth]{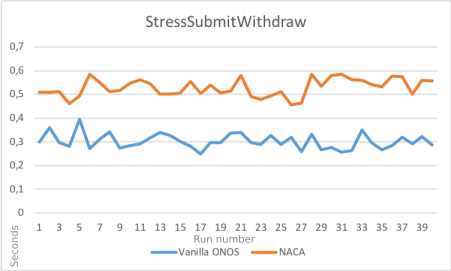}
	\caption{\texttt{StressSubmitWithdraw}, time in s.}
	\label{fig:submitWithdraw}
\end{figure}

Finally, Figure~\ref{fig:submitWithdraw} illustrates an evaluation of a combined stress test where intents are submitted and subsequetly withdrawn.

Similar to intent compilation, we believe the performance of intent submission and intent withdrawal can be improved by implementing the reference monitor in an isolated execution environment on the same host.
This will reduce the communication overhead both between the network controller and the reference monitor, and between the request tagger and the network controller.
A similar strategy or decomposing elements in the data plane was sucessfully demonstrated in~\cite{paladi:2017d}, \cite{paladi:2018}
Detailed statistics are listed in Table~\ref{tab:ovh-stats}.

\begin{table}[h]
 \scriptsize
\begin{center}
\caption{Summary of performance evaluation of $\mathsf{NACA}$}
\label{tab:ovh-stats}
  \begin{tabulary}{\textwidth}{|L | C| C | C | C | C |}
  	\hline
    \textit{Data} 									    & \textit{Minimum} & \textit{Maximum} & \textit{Mean} & \textit{Median} & \textit{Stddev} \\
    \hline
	compileIntent, s           								& 0,154			    &0,287		 &0,192	    		 &0,1855			   		& 0,033        \\
    \hline
	with $\mathsf{NACA}$								& 0,16				&0,298				 &0,207			&0,203					& 0,0314  \\	
    \hline
    Overhead												&		 		   	   &					&\textbf{7,89\%}		&\textbf{9,43\%}			    &		\\
    \hline
    \hline
	submitIntent, s           							& 0,268			    &0,488		 &0,323	    		 &0,306			   		& 0,057           \\
    \hline
	with $\mathsf{NACA}$											& 0,467				&0,681				 &0,537			&0,526			& 0,526  \\	
    \hline
    Overhead												&		 		   	   &					&\textbf{66\%}		&\textbf{72\%}			    &		\\
    \hline
    \hline
	stressSubmitWithdraw, s 								&0,250			    &0,396			  &0,302			&0,297				& 0,032  \\
    \hline
	with $\mathsf{NACA}$ 								&0,460					&0,762			 &0,579			&0,572				& 0,063  \\
    \hline
    Overhead												&		 		   &					& \textbf{72\%}		&\textbf{72\%}		    &		\\    
    \hline
  \end{tabulary}  
\end{center}
\end{table}

We did not evaluate \textit{multi-application scalability} in this context. 
This work primarily focuses on exploring the principle of making high-level network configuration and control commands (intents) traceable (through tagging) and enforceable (through access masks) throughout the evaluation and execution flow.
However, we intend to further extend this work and evaluate multi-application scalability in an extended prototype of $\mathsf{NACA}$.


\section{Related Work}
\label{sec:related_work}
From the origins of SDN, access control and network programambility have received significant attention.
\paragraph{Access control for SDN controllers}
Casado et al. describe in Ethane~\cite{casado:2007} an enterprise network architecture allowing network managers to control deployments using system-wide fine-grained policies.
Ethane includes two component types: 
\textit{(1)} `dumb' switches maintaining flow table entries and communicating with the controller,
and \textit{(2)} one or more controllers handling host registration and authentication, tracking network bindings, implementing access control, and enforcing resource limits on the managed flows. 
The proposed high-level language for Ethane network management policies -- despite its shortcomings such as lack of support for dynamic policy updates and assumption of a fixed network topology -- has inspired
a rich collection of subsequent network policy languages~\cite{foster:2011,voellmy:2012,monsanto:2013,mcclurg:2016}.
While such network control languages fueled a rapid development of network controller capabilities, they operate on a lower abstraction level than required for network management applications.
Several outstanding issues of the Ethane model are broadcast and service discovery, application-layer routing, knowledge about application-layer configuration and potential damage from spoofing Ethernet addresses.

Ferguson et al.~\cite{ferguson:2012} rely on \textit{hierarchical} composition of policies to define access to actions performed on traffic flows, as well as to control resource allocation.
Furthermore, the framework contains a policy conflict resolution mechanism based on user-defined operators.
Conflict resolution is essential to enable a distributed approach to policy definition, where distinct applications requiring device resources and services define access policies and Quality-of-Service requirements.
While this approach brings control over network access to applications using network services, it does not hide the details of device resources and requires from the application knowledge of the network internals.
$\mathsf{NACA}$ maintains the higher-level control of network access, while abstracting the network internals from the application programmers.

The access control scheme for SDN controllers proposed in \cite{klaedtke:2014} accounts for device resources, multiple security requirements, conflicts originating from reconfiguration of network components, and delegation of access permissions.
This scheme mimics access control schemes for operating systems, with contextual adjustments.
In this case network users are the subjects and the network components are the objects.
A follow-up prototype is described in~\cite{gkounis:2016}.

\paragraph{APIs for Network Configuration}
The \textit{Pane} controller~\cite{ferguson:2013} allows applications to control SDN deployments.
It delegates read-write privileges on network configuration to end-users or applications acting on their behalf.
The \textit{Pane} API allows three messages types -- requests, queries and hints. 
Requests affect the state of the network following the intention of the application.
Applications can issue queries about the state of the network and provide hints to notify the controller about upcoming changes, e.g. in the network resource usage.
In this model, principals are included in one or more shares and have complete access control privileges -- including recursive delegation of privileges -- over the set of network flows in the respective shares.
Intra-share resource over-subscription is allowed and is monitored by the controller, which also detects and resolves potential conflicts.
$\mathsf{NACA}$ takes a different approach by restricting applications to network control privileges explicitly declared, verified and validated at deployment time.
Furthermore, it verifies that control messages issued by network applications comply with the allowed control privileges.
 
A low-level API for dynamic configuration of Quality-of-Service resources in network devices can be implemented through a plugin to the OVSDB protocol~\cite{caba:2015}.
Such low-level configuration allows granular control across three levels: port, switch and network-wide controls.
While this allows extensive control over the forwarding plane configuration, the proposed approach does not discuss mechanisms to prevent applications using such an API from monopolizing network control, nor any resolution mechanism in case of conflicting policies.

A Network Overlay Framework~\cite{trois:2015} addresses the challenge of configuring network deployments according to application requirements.
Through its API and programming language, the Network Overlay Framework allows developers to program the network according to the needs of the application.
This replaces the ``best-effort" packet delivery approach with explicit Quality-of-Service guarantees tailored to a specific application.
While the approach has been implemented for the Hadoop data processing framework, it does not consider any multi-application or multi-tennant scenarios.
Likewise, it does not describe any resolution mechanism that would be needed in the case of configuration conflicts. 
Furthermore, this approach allows the forwarding plane to manipulate the configuration of the control plane, disregarding the possibility of malicious or misconfigured applications. 
This in turn creates network security and safety concerns.

SDN controller functionality can be implemented by extending an existing operating system and leveraging its software ecosystem, i.e. operating system utilities and a distributed file system, as proposed in \textit{yanc}~\cite{monaco:2013}.
Here, file I/O is used as a single API for SDN applications, allowing to avoid the restriction to use programming languages mandated by the implementation of the SDN controller.
The authors indicate the possibility of using permissions implemented by the virtual file system layer for fine-grained access control of network resources.
\textit{Yanc} could leverage $\mathsf{NACA}$ by storing \textit{resource masks} in the extended file attributes and further integrating the $\mathsf{NACA}$ API with existing access control policies.

Eden is a framework for enabling end-host network functions~\cite{ballani:2015}, assuming a single-domain network deployment, where at least a subset of end-hosts can be trusted, e.g. in datacenters.
The design builds on the observation that a large class of network functions feature three key requirements: data-plane computation, data-plane state, and operation on application semantics.
It includes a flexible scheme for application-level classification of network traffic based on a custom language as well as a compiler and run-time for action functions -- constructs able to access and modify packet classes and endpoint enclaves.
In Eden, network applications are ``first-class principals", and can classify packets based on application-internal semantics. 
The classification follows the packet through the end host stack and is used to determine the rules to apply.
A similar approach has been used in $\mathsf{NACA}$ for access control.
However, Eden assumes a single-domain, controlled and trusted environment, and does not feature policy conflict resolution or access control mechanisms.

An early concept of intent-based network abstractions has been introduced in~\cite{cohen:2013}, allowing to specify networks as a policy governed service.
Such intents describe the functionality of the network -- i.e. \textit{what} connectivity a certain application requires, connectivity between endpoints and policies associated with the connectivity -- while leaving out the specifics on \textit{how} to implement the required connectivity.
This approach is further developed in projects such as ``Boulder North-bound Interface (NBI)''~\cite{nbi:2016}.
$\mathsf{NACA}$ reuses the NBI implementation in the ONOS intent framework and extends the intent concept described in~\cite{cohen:2013} with an API to specify access control constraints on the intents.

Maple~\cite{voellmy:2013} enables developers to use programming languages such as Haskell and Java to define network behaviors through centralized algorithms (algorithmic policies).
The use of algorithmic policies hides the challenges of implementing high-level policies into sets of rules on distributed individual switches.

In large distributed systems such as SDN deployments configuration updates may lead to undefined network behavior and security vulnerabilities, if applied incorrectly due to long latency, dropped packets or weak consistency.
To address this, \textit{event-driven consistent updates}~\cite{mcclurg:2016} preserve well-defined behaviors when transitioning between configurations in response to events. 
The approach places strong \textit{locality} requirements towards configuration updates: it allows exclusively configuration changes decidable with \textit{local} (rather than remote) information in a distributed system.
$\mathsf{NACA}$ allows to control the locality access of network applications to SDN resources (recall \textit{Extent} and \textit{Placement} resource masks in Table~\ref{tab:resource-mask-attributes}), and can be extended to include other access types.

\section{Limitations and Future Work}
\label{sec:future_work}
The $\mathsf{NACA}$ mechanisms is only a first step towards implementing usable and scalable deployment and configuration mechanisms for software defined network deployment.
As such, it must be further improved and refined to reach wider adoption.
Therefore, in future work we aim to extend and improve several aspects of this contribution, as follows:
first, improve the performance of $\mathsf{NACA}$;
second, deployment of dedicated components may be seen as a drawback in certain contexts.
Re-implementing the request tagger and reference monitor as components of the network controller running in an isolated execution environment will help address this use case, as well as potentially improve the performance of $\mathsf{NACA}$.
Third, integrate with other controller platforms, in particular \textit{yanc}~\cite{monaco:2013} in order to adapt the $\mathsf{NACA}$ intent approach to existing operating system access control mechanisms.
Fourth, extend $\mathsf{NACA}$ with \textit{hints} as in~\cite{ferguson:2012} can improve the overall performance of the deployment.
Finally, implement and evaluate the approach using alternative isolation solutions discussed in~\S\ref{subsec:compis}, to generate new insights into trade-offs between security guarantees and performance for network controllers.

\section{Conclusion}
\label{sec:conclusion}
Design and implementation of a controller-agnostic north-bound interface is the current focus of SDN development.
Such an interface will allow network operators to deploy multiple applications on their SDN infrastructure and unlock rich network management functionality.
However, the currently available north-bound interfaces do not offer access control mechanisms to allow applications to negotiate network resource access
or operators to obtain a comprehensive overview of the resource access of the installed applications.
We have addressed this by first introducing a taxonomy of resource access models for SDN infrastructure, along with a network access control API which allow applications to commit at deployment time to a set of resource access requirements, which are then enforced by discrete components on the network controller platform.
We described the design, implementation and performance evaluation of the proposed solution.

\section{Acknowledgments}
The research was conducted within the COLA project and received funding from the European Union's Horizon 2020 research and innovation programme under grant No 731574.

\bibliographystyle{spphys}       
\bibliography{aac}   

\end{document}